# ASSESSMENT ON LSPU-SPCC STUDENTS' READINESS TOWARDS M-LEARNING

Joanna A. Erlano-De Torres

College of Computer Studies,
Laguna State Polytechnic University-San Pablo City Campus, Philippines

## ABSTRACT

*Today, the use of technology is a powerful advantage in every field in the society. With the advent of development in information and communications technology (ICT), the process of learning and acquiring new knowledge had undergone a shift marked by a transition from desktop computing to the widespread use of mobile technology. In light of the COVID-19 pandemic, the Commission on Higher Education said that colleges and universities following the new school calendar will no longer require students to attend face-to-face classes. One of the state universities that had been affected by this inevitable situation is the Laguna State Polytechnic University. This study aims to determine the readiness of the students in shifting to m-learning. Specifically, it aims to determine the availability of mobile devices, equipment readiness, technological skills readiness and psychological readiness. A survey-based methodology was used to obtain the data and descriptive statistics to analyze the results. It was determined that almost all of the students own mobile devices, are fully equipped with applications, have high technological skills and are quite ready in terms of psychological readiness.*

## KEYWORDS

## 1. INTRODUCTION

Today, the use of technology is a powerful advantage in every field in the society. With the advent of development in information and communications technology (ICT), the process of learning and acquiring new knowledge had undergone a shift marked by a transition from desktop computing to the widespread use of mobile technology. Technology is changing the way people learn, work, conduct business, access information, and interact [1]. According to Almutairy, Davies, & Dimitriadi [2], the learning process no longer relies on traditional teaching methods but has instead expanded to include new technologies and forms of learning; such as mobile learning (m-learning). Mobile technologies offer diverse opportunities to deliver innovative and interesting modes of learning, both inside and outside the classroom.

M-learning is an educational model that emerged along with the development of mobile technologies [3]. This kind of technology requires the utilization of Information and Communication Technology to accomplish the learning process. As Abas, Peng & Mansor [4] stated, it offers enormous potential as a tool to be used in situations where learners are geographically dispersed, to promote collaborative learning, to engage learners with content, as an alternative to books or computers, as an alternative to attending campus lectures and for 'just-in-time' delivery of information.

The statistics shows that in 2019, 44.3 million Filipino people accessed the internet through their mobile phones, and in the year 2023, this figure is projected to amount to 50.8 million mobile phone internet users [5]. Thus, learning is no more restricted to classroom settings and conducted by the instructors only. Rather, it has reached a horizon that involves the application of portable devices as well as wireless technologies and allows the learners to learn anywhere and anytime [6].

Currently, there is a global pandemic called coronavirus disease 19 (COVID-19) is a highly transmittable and pathogenic viral infection caused by severe acute respiratory syndrome coronavirus 2 (SARS-CoV-2), which emerged in Wuhan, China and spread around the world. According to Shereen, Khan, Kazmi, Bashir & Siddique [7], the genomic analysis revealed that SARS-CoV-2 is phylogenetically related to severe acute respiratory syndrome-like (SARS-like) bat viruses, therefore bats could be the possible primary reservoir. The intermediate source of origin and transfer to humans is not known, however, the rapid human to human transfer has been confirmed widely. There is no clinically approved antiviral drug or vaccine available to be used against COVID-19.

In light of this pandemic, the Commission on Higher Education said that colleges and universities following the new school calendar will no longer require students to attend face-to-face classes. One of the state universities that had been affected by this inevitable situation is the Laguna State Polytechnic University. This academic institution consists of four strategically-placed campuses in Sta Cruz, Los Banos, Siniloan, and San Pablo City. In San Pablo City campus alone, there are 7,636 students from senior high school up to graduate studies to be catered. Instructors and Professors were advised to provide learning modules to the students online to be able to compensate for the weeks, even months of learning opportunities that had been lost. Hence, m-learning readiness of Laguna State Polytechnic University-San Pablo City Campus must be assessed.

## 1.1. Objectives

The main objective of the research project is to assess the readiness of the undergraduate students of Laguna State Polytechnic University-San Pablo City Campus towards mobile learning or m-learning.

Specifically, it aims to:

1. Determine if all students have access to a mobile device;
2. Determine the readiness of equipment in terms of messaging applications, social media applications, making video calls, connecting to wireless fidelity (Wi-Fi), subscribing to data plans, ability to open word, PDF, Excel, PowerPoint, video files, audio files, and photographic files, ability to edit video, audio and photo, resolution of the camera (> 5 megapixels), and storage capacity;
3. Determine the students' technological skills readiness; and
4. Determine the students' psychological readiness.

## 2. RELATED LITERATURE

### 2.1. Mobile Devices among Students

The evolution of handheld portable devices and wireless technology has resulted in radical changes in the people's lifestyles around the world, including for learning [8]. Students' personal mobile devices can potentially enhance classroom learning experience [9]. It can also encourage students to learn anytime and anywhere because the students can process information inside and outside the classroom.

Findings from studies stated that mobile devices allowed students to conduct nine (9) activities in higher education as the following: a) to send pictures or movies to colleagues, b) to use mobile phone as MP3 player, c) to access information or services on the web, d) to make video calls, e) to take digital photos or movies, f) to send or receive email, g) to use mobile phone as a personal organizer (e.g. diary, address book), h)to send or receive SMS to colleagues, and i) to call the colleagues or others [10],[11].

## 2.2. M-learning

Mobile learning or m-learning is a rising art of using mobile technologies to enhance the learning experience [12]. As been highlighted in [13], mobile technologies can significantly reduce people's dependence on fixed locations, and thus have the potential to revolutionize the way people work and learn. In fact, some other studies also indicate the potential of mobile technologies in assisting the teaching and learning process in school [14], [15].

By definition, mobile learning (m-learning) is learning through wireless technological devices that can be pocketed and utilized wherever the learner's device can receive unbroken transmission signals [16]. As Nassoura [17] pointed out, "…M-learning provides an opportunity for the new generation of people with better communication and activities without taking into account the place and time".

Moreover, Al-Said [18] pointed out that using portable mobile devices in teaching, learning, and training provide the learners and trainees the ability to access the learning materials continuously, anytime anywhere, and at the same time, provide the teachers and trainers the ability to easily deliver homework activities continuously without interruption for learners and trainees, and that are parts of the educational process, which may not be provided by e-learning. M-learning and Edmodo applications can take place everywhere, every time at home, in a car, day, night, etc. since mobile devices are lighter and less bulky from bags full of books, papers, or even laptops.

Implementing M-learning requires a high level of commitment from both lecturers and students; otherwise, it would neither be feasible nor effective. Accessing the internet, sending and replying SMS involve certain expenses. Besides that, it is also important to ascertain students' technology readiness before implementing M-learning [19].

According to Chaka and Govender [20], some studies have argued that m-learning is an extension of e-learning, but that it differs in the sense that it uses mobile devices rather than computers as a medium [21,22]. Park [23] attributes the increasing popularity of mobile learning to new innovations in application and social networking sites including wikis, blogs, twitter, and Facebook among others.

## 2.3. Technology Readiness

Technology readiness (TR) is defined as the propensity to embrace and use new technologies for accomplishing goals in home life and at work [24]. The TR concept is widespread, particularly in the business marketing domain where research focuses on identifying segments of the market who are likely to adopt new technologies such as mobile data services [25], and online insurance [26], among others. In each of these studies, the authors found the technology readiness model to be effective for studying respondents' propensity to adopt new technologies [27].

Moreover, the extent to which individuals desire to use new technology is commonly influenced by such factors as culture [28, 29], attitudes toward specific technologies [30, 31], the level of technology anxiety exhibited by individuals [32] and an individual's capacity and willingness to use [33].

# 3. METHODOLOGY

This is a survey study using quantitative methods that seeks to explore students' readiness in m-learning. A survey-based methodology is commonly used to analyze the results obtained from questionnaires. This study acquired data for two weeks in September 2019 and the last week of April 2020, using an online survey (Google Form) consisting of 50 questions. The data collection was administered to 295 students following the random sampling technique. The questionnaire was divided into five sections, namely: (1) demographic data, (2) about mobile devices, (3) equipment readiness (4) technological skill readiness, and (5) psychological readiness. A five-point Likert scale was used to measure responses to the main questions, with ratings of 'strongly disagree', 'disagree', 'neutral', 'agree' and 'strongly agree'. Because of the nature of the quantitative data obtained by the survey, the common analysis method used in this study was descriptive statistics which included the numbers of frequencies and percentages. The analysis was achieved through SPSS software. Data obtained assisted the researcher to answer the study questions.

## 3.1. Demographic Information

The respondents in this research are undergraduate students studying at Laguna State Polytechnic University-San Pablo City Campus (LSPU-SPCC). The sample comprised students of both genders, aged 16 to 28 years. The quality of a piece of research stands or falls not only by the appropriateness of methodology and instrumentation but also by the suitability of the sampling strategy adopted.

The sample, therefore, was selected randomly from different colleges and students at different levels of study, to cover a diverse group of LSPU-SPCC students using mobile devices for learning purposes.

Table 1. Students' Demographic Information

| ITEMS | N=295 | | |
|---|---|---|---|
| | *Frequency* | *%* | *Cumulative* |
| **1. Age** | | | |
| **16-17** | 5 | 1.69% | 5 |
| **18-21** | 280 | 94.92% | 285 |
| **>22** | 10 | 3.39% | 295 |
| **2. Sex** | | | |
| **Male** | 181 | 61.36% | 181 |
| **Female** | 114 | 38.64% | 295 |
| **3. Highest Educational Attainment** | | | |
| **High School** | 74 | 25.08% | 74 |
| **College** | 221 | 74.92% | 295 |
| **4. Location** | | | |
| **Urban Setting** | 57 | 19.32% | 57 |
| **Rural Setting** | 238 | 80.68% | 295 |

Table 1 presents the distribution of the study sample according to demographic variables. The data reveals that the majority of the students responded were 18-21 years old (94.92%). It also reveals that out of 295 students, 38.64% were female and 61.36% were male. As for the highest educational attainment, 74.92% were in college. In addition, most were educated to college level (74.92%). Lastly, 80.68% of the students live in a rural setting, meaning, they live outside the densely populated urban areas in a town or city. Rural areas are traditionally areas with large, open areas with few houses and few people, as opposed to urban areas which have larger populations.

## 4. RESULTS AND DISCUSSIONS

The discussion for these research findings is based on the factors studied- availability of mobile devices, readiness of the device (s), students' technological skills, and students' psychological readiness.

### 4.1. Availability of Mobile Device(s)

Mobile devices have progressed from convenience in daily life to a necessity. In higher education, the use of mobile technologies in learning has also increased rapidly over the years.

Table 2. Availability of Mobile Device(s)

| ITEMS | N=295 | | |
|---|---|---|---|
| | *Frequency* | *%* | *Cumulative* |
| 1. Do you own any mobile devices (phone, tablet, ipad etc.) | | | |
| Yes | 291 | 98.64% | 291 |
| No | 4 | 1.36% | 295 |
| 2. Do you carry your mobile devices with you? | | | |
| Always | 243 | 82.37% | 243 |
| Usually | 36 | 12.20% | 279 |
| Sometimes | 12 | 4.07% | 291 |
| Never | 4 | 1.36% | 295 |
| 3. How many mobile devices you usually bring with you? | | | |
| >3 | 7 | 2.37% | 7 |
| 2 | 37 | 12.54% | 44 |
| 1 | 247 | 83.73% | 291 |
| 0 | 4 | 1.36% | 295 |

Table 2 reveals that 98.64% of the respondents own mobile devices such as phone, tablet, and iPad. The majority of the students responded that they always bring their mobile devices (82.37%), 12.20% usually bring them, 4.07% sometimes bring them, and 1.36% never bring one since they don't have it. Overall, the majority of the students have access to basic ICT equipment (mobile devices as a type of personal computer) that can help them in utilizing m-learning opportunities.

## 4.2. Students' Equipment Readiness

Equipment in this study pertains to the mobile devices that the students own. Determining if the equipment is capable of engaging in different m-learning activities is essential.

Table 3. Students' Equipment Readiness

| ITEMS | N=295 | | |
|---|---|---|---|
| | *Frequency* | *%* | *Cumulative* |
| 1.Do you have messaging app (messenger, wechat, viber etc.) on phone? | | | |
| Yes | 291 | 98.64% | 291 |
| No | 4 | 1.36% | 295 |
| 2. Do you have a social media app (FACEBOOK, INSTAGRAM, twitter etc.) on phone? | | | |
| Yes | 293 | 99.32% | 293 |
| No | 2 | 0.68% | 295 |
| 3.Can your mobile device make video call? | | | |
| Yes | 284 | 96.27% | 284 |
| No | 11 | 3.73% | 295 |
| 4. Can your mobile device connect to WiFi? | | | |
| Yes | 291 | 98.64% | 291 |
| No | 4 | 1.36% | 295 |
| 5. Do you subscribe to any Data Plan? | | | |
| Yes | 113 | 38.31% | 113 |
| No | 182 | 61.69% | 295 |
| 6.Can your mobile device has the ability to open word document? | | | |
| Yes | 278 | 94.24% | 278 |
| No | 17 | 5.76% | 295 |
| 7. Can your mobile device has the ability to open PDF document? | | | |
| Yes | 270 | 91.53% | 270 |
| No | 25 | 8.47% | 295 |
| 8. Can your mobile device has the ability to open excel document? | | | |
| Yes | 249 | 84.41% | 249 |
| No | 46 | 15.59% | 295 |
| 9. Can your mobile device has the ability to open PowerPoint document? | | | |
| Yes | 272 | 92.20% | 272 |
| No | 23 | 7.80% | 295 |
| 10. Can your mobile device has the ability to open video files? | | | |
| Yes | 290 | 98.31% | 290 |
| No | 5 | 1.69% | 295 |
| 11. Does your mobile device has the ability to open photo/graphic files? | | | |
| Yes | 281 | 95.25% | 281 |
| No | 14 | 4.75% | 295 |
| 12. Does your mobile device has a video editing app? | | | |
| Yes | 149 | 50.51% | 149 |
| No | 146 | 49.49% | 295 |
| 13. Does your mobile device has a photo editing app? | | | |
| Yes | 213 | 72.20% | 213 |
| No | 82 | 27.80% | 295 |
| 14. Is you hand phone's camera more than 5 megapixel? | | | |
| Yes | 251 | 85.08% | 251 |
| No | 44 | 14.92% | 295 |
| 15. Memory capacity of your device? | | | |
| >128 | 17 | 5.76% | 17 |
| 64 | 40 | 13.56% | 57 |
| 32 | 87 | 29.49% | 144 |
| 16 | 77 | 26.10% | 221 |
| 8 | 36 | 12.20% | 257 |
| <8 | 38 | 12.88% | 295 |

Table 3 reveals that 98.64% of the mobile devices that the students own has messaging applications such as Facebook messenger, WeChat, and Viber. As for the social media applications, 99.32% of the respondents stated that they had installed Facebook, Instagram, Youtube, SnapChat, TikTok and similar applications. In addition to this, 96.27% stated that their mobile devices can make video calls. For connectivity, 98.64% stated that their mobile devices can connect to a Wi-Fi while 38.31% have data plan subscriptions. For opening word documents, 94.24% specified that their mobile devices are capable of doing that. As for opening PDF file, 91.53% answered that their mobile devices are also capable. For a more complicated application such as Excel, 84.41% revealed that their mobile devices can open it. In addition, 92. 20 % of the respondents stated PowerPoint application can be opened on their mobile devices. Also, 98.31% of them answered that their mobile devices can open video files while 95.25% stated that their mobile devices can open photo/graphic files. The mobile devices that the 50.51% of respondents have video editing applications while 72.20 % has photo editing applications. As for the mobile devices' camera resolution, the majority of the respondents (85.08%) have more than 5 megapixels. Lastly, the highest storage capacity of their mobile devices is 32 gigabytes (29.49%) followed by 16 gigabytes (26.10%), 64 gigabytes (13.56%), less than 8 gigabytes (12.88%), 8 gigabytes (12.20%) and more that 128 gigabytes (5.76%).

Overall, the table above indicated that respondents' equipment (mobile devices) can support m-learning since their mobile devices have essential applications to open and edit learning materials.

### 4.3. Students' Technological Skills Readiness

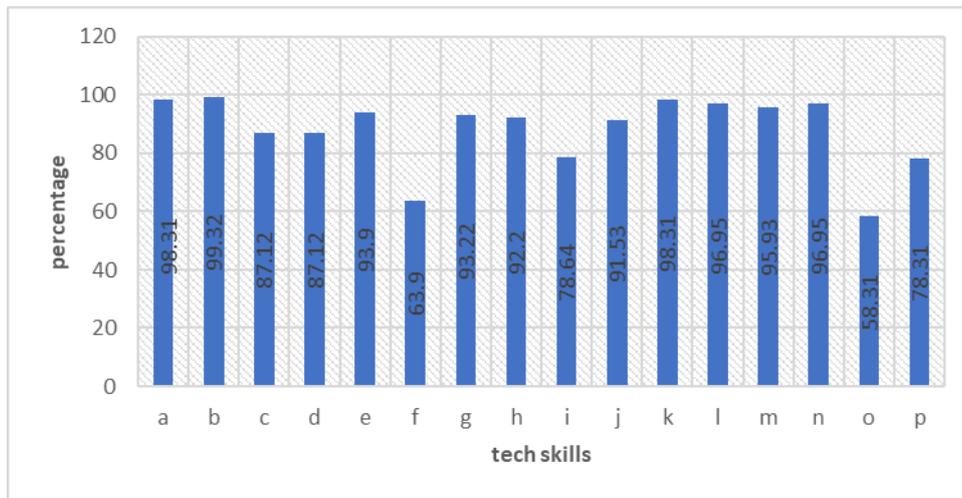

Figure 1. Technological Skills Readiness of the Students

The figure 1 reveals that in the aspect of technological skills among students, there were 16 items such as (a) 98.31% of the students know how to properly use messaging applications such as Messenger, WeChat, and Viber on mobile devices, (b) 99.32% of the students own accounts in social media applications like Facebook, Instagram, YouTube, and Twitter, (c) 87.12% of the students know how to use video call services on their mobile devices, (d) 87.12% of the students know how to send and receive emails on their mobile devices, (e) 93.9% of the students know how to use WiFi to connect to the internet using their mobile devices, (f) 63.9% of the students know how to utilize data plan to connect to the internet using a mobile device, (g) 93.22% of the students know how to open/ read word documents on mobile devices, (h) 92.2% of the students know how to open/ read pdf documents on mobile devices, (i) 78.64% of the students know how to open/ read excel document on their mobile devices, (j) 91.53% of the students know how to open/ read PowerPoint documents on their mobile devices, (k) 98.31% of the students know how to view video files on mobile devices, (l) 96.95% of the students know how to listen to audio files on mobile devices, (m) 95.93% of the students know how to open/view photo or graphic files on mobile devices, (n) 96.95% of the students know how to download files (document, audio, video, photo) on mobile devices,(o) 58.31% of the students know how to edit video using mobile devices and (p) 78.31% of the students know how to edit photos using mobile devices. Overall, students have the technical know-hows in maximizing their mobile devices' full potential.

### 4.4. Students' Psychological Readiness

Psychological Readiness of the students pertains to the confidence in one's ability to perform; one's fears need to be under-control. This does not mean that the students cannot have fears. It means their fears do not have the power to change what their bodies are physically able to do.

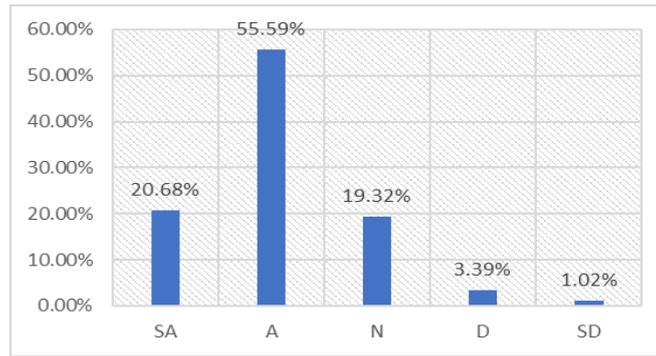

Figure 2. I like to use my own mobile phone in my own learning processes

This figure shows that the majority of the students (55.59%) agreed to the statement that they like to process their own learnings using their mobile phones. Only 1.02% strongly disagreed with the idea. This only means that they can study their lessons using their mobile phone since some of the teachers opted to upload their instructional materials on social media like Facebook.

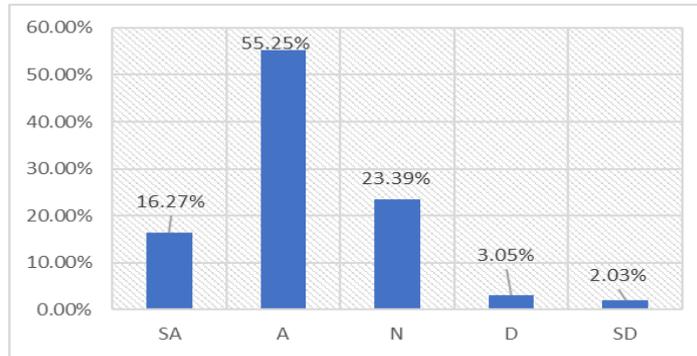

Figure 3. I am confident when using my mobile phone for learning.

Figure 3 shows that more than half of the student respondents (55.25%) agreed to the statements that they are confident when using their mobile phones for learning. Only 2.03% strongly disagreed with the idea. This only means that learning is possible even without other ICT equipment like computers because they are confident in utilizing their mobile phones to do some of their school work.

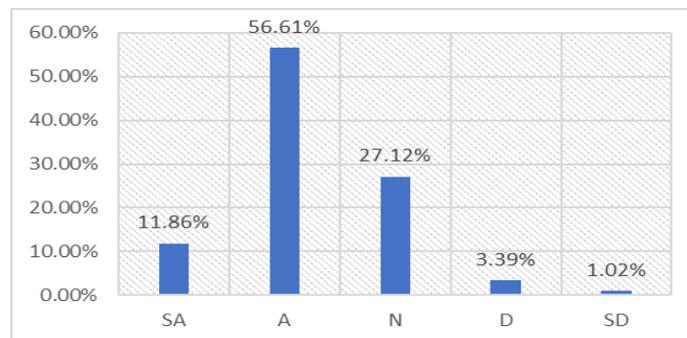

Figure 4. M-learning provides me with new methods to learn.

Figure 4 shows that 56.61% students agreed that m-learning can provide them with new methods to learn. Only 1.02% strongly disagreed with the idea. This outlook is a positive sign that students want to evolve in terms of acquiring knowledge other than traditional face to face lectures.

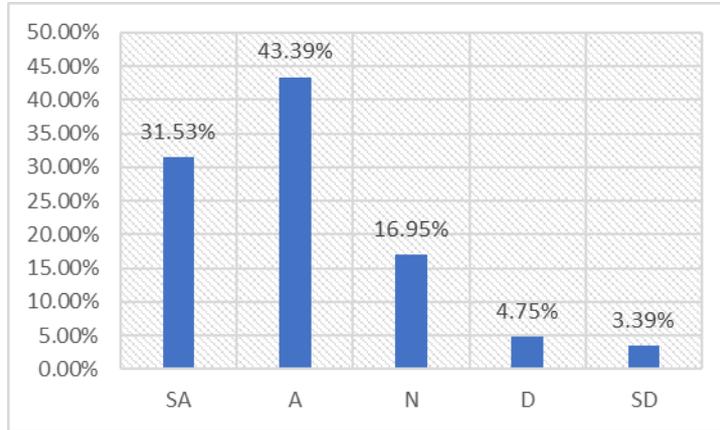

Figure 5. I would be interested in owning a new mobile device with advanced features if it would improve my learning and performance at university

Figure 5 shows that three quarters of the student respondents (31.53% strongly agreed and 43.39% agreed) were interested in owning a new mobile device with advanced features if it would improve my learning and performance at university. This only shows that students have keen interest in learning and improving themselves academically. Only 3.39% strongly disagreed because they believed that their present mobile devices can do the job.

As for figure 6, 19.32% strongly disagreed while a majority of 56.95% agreed that m-learning will bring new opportunities for learning because they can access the internet, their main source of information, anytime and anywhere using WiFi or data plans. Only 1.36% strongly disagreed since they do believe that school is still the best opportunity provider when it comes to learning.

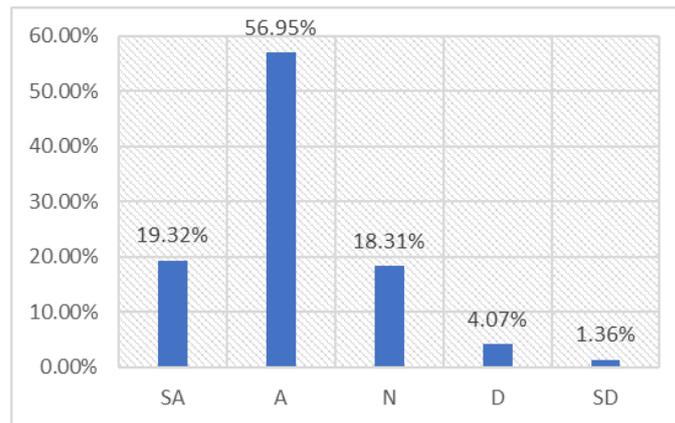

Figure 6. M-learning will bring new opportunities for learning

In Figure 7, the majority of the student respondents having 16.27% strongly agreed and 48.81% agreed that m-learning can save time. This response can be contributed to the fact that students can easily access online learning materials when needed, using their mobile phones. Only 4.41% strongly disagreed because they do believe that using mobile phones in doing school tasks can sometimes lead them to distractions like going to their social media accounts instead of reading instructional materials.

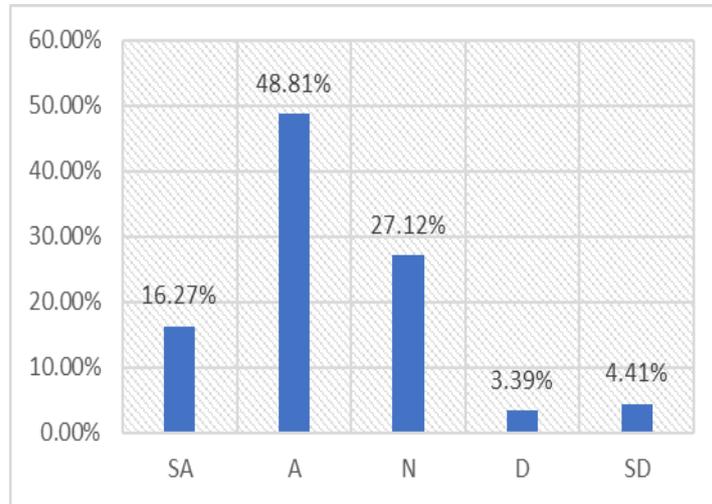

Figure 7. M-Learning can save my time

In figure 8, the majority of the student respondents (60.68%) find m-learning easy, as it is possible to learn what they want to learn. On the other hand, only 1.02% strongly disagreed because they believe that it is better to learn the traditional way, and they don't have the means to subscribe to internet service providers and data plans.

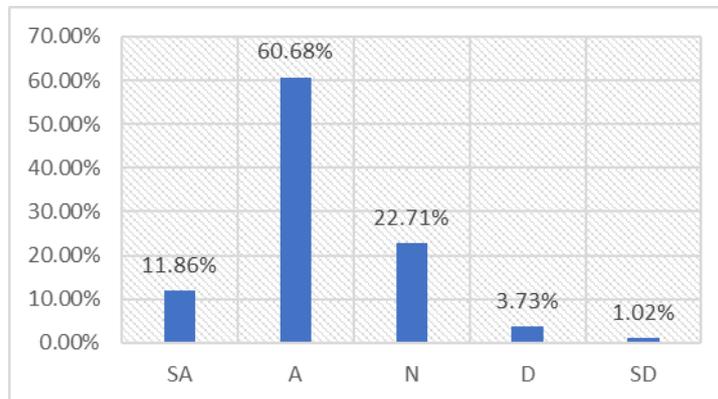

Figure 8. I find m-learning easy, as it is possible to learn what I want.

As for figure 9, most of the respondents (61.02%) agreed to the statement that m-learning meets their needs and interests. Students, nowadays, want instant results when it comes to researching school assignments and projects as well as they favor using mobile devices as their learning tool rather than their notebooks and lecture handouts. On the other hand, 2.03% strongly disagreed since they prefer writing and reading their own study guides.

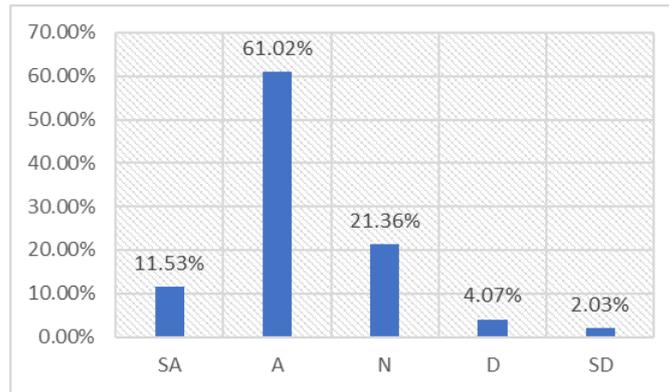

Figure 9. M-learning meets my needs and interests

In figure 10, more than half of the student respondents (54.92%) agreed while 12.20% strongly agreed that m-learning enables them to get feedback from lecturers more quickly than before. They are using the social media platform where their teachers post their lectures to communicate with them if they need to clarify or be enlightened on some topics. Only 0.68% strongly disagreed to this statement for unknown reasons.

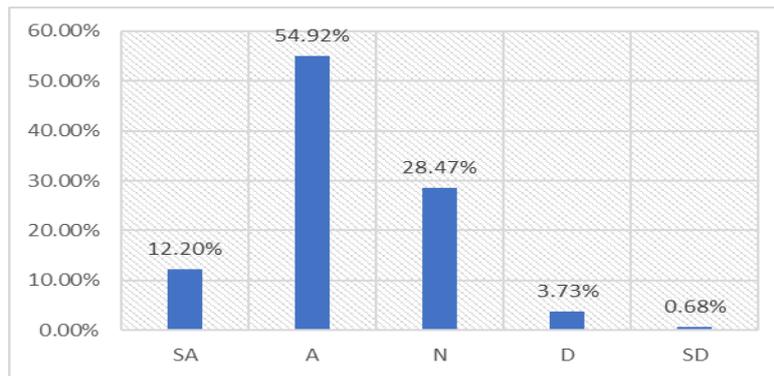

Figure 10. M-learning enables me to get feedback from lecturers more quickly than before.

In figure 11, 14.92% of the respondents strongly agreed and 55.25% agreed that m-learning is more flexible than traditional learning. It can be carried out at any time, and anywhere. Only 1.69% strongly disagreed with it for the reason that they need more time to adjust to this kind of learning.

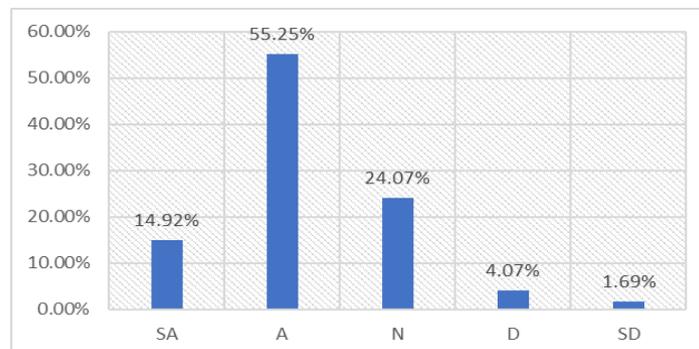

Figure 11. M-learning is flexible than traditional learning

In figure 12, the result shows that more than half of the respondents (54.24%) agreed to the statement that it is possible to achieve personal educational aims through m-learning. It is believed to be beneficial for doing school activities because they can read the lessons in advance and they can review previous lessons, also. As for the 1.69% of the respondents, they strongly disagreed with the statement because for them, it is better to study in school rather than at home because they can focus more.

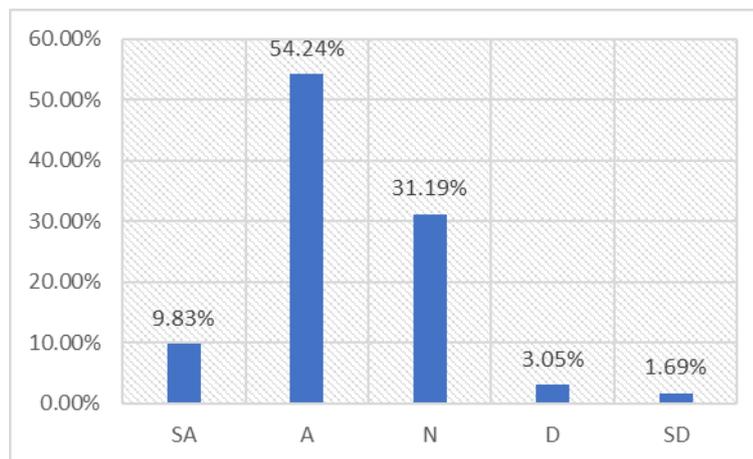

Figure 12. It is possible to achieve personal educational aims through m-learning.

In figure 13, the result shows that 20.34% strongly agreed and 48.47% agreed that they would be happy if they could use their mobile phone in the classroom to support their learnings. Only 3.05% strongly disagreed because for them, using mobile phones while the teacher is discussing can be both a distraction and a lack of etiquette.

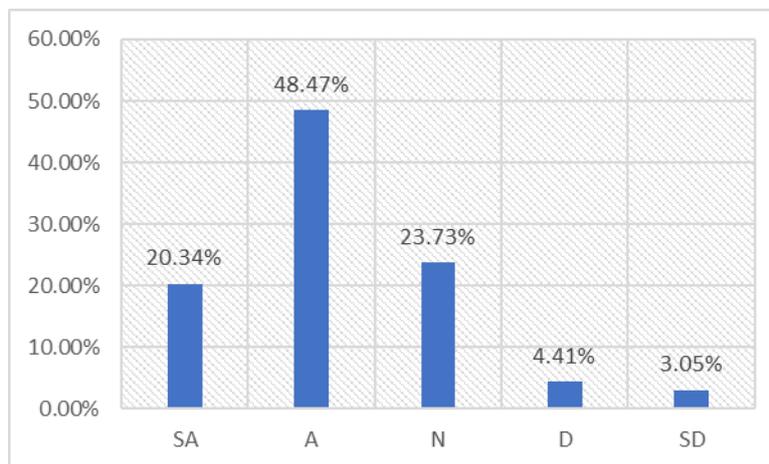

Figure 13. I would be happy if I could use my mobile phone in the classroom to support my learning.

In figure 14, majority of the respondents agreed to the statement that the university is ready for mobile learning using mobile phone facilities because they can see that there is strong support from the school administration in terms of providing ICT services. Last year, one of the projects of the university, together with Senator Bam Aquino, to become fully equipped with wireless fidelity came into reality. Students freely enjoyed free WiFi to perform online tasks and assignments. On the other hand, 3.05% strongly disagreed with the statement.

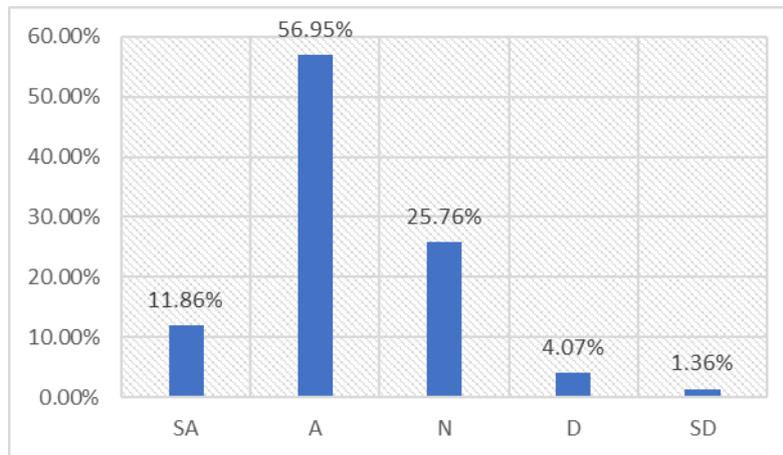

Figure 14. I think my university is ready for mobile learning using mobile phone facility

In figure 15, 15.25% strongly agreed and 54.58% agreed that some of the lecturers are now integrating m-learning in teaching. This is evident in FB groups and Google classroom to name a few. On the other hand, 3.39% strongly disagreed because they seldom check their online accounts in social media.

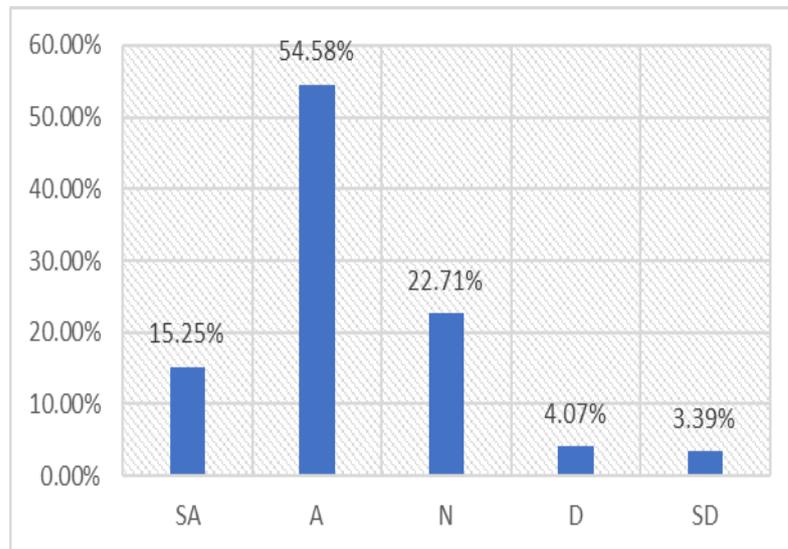

Figure 15. Some of the lecturers are integrating m-learning in their teaching

In figure 16, more than half of the students (56.95%) agreed that m- learning is alternative to web-based learning since their mobile phones can access different e-learning platforms. However, 1.36% strongly disagreed with the statement because they don't feel confident in the idea of putting too much work using their mobile devices.

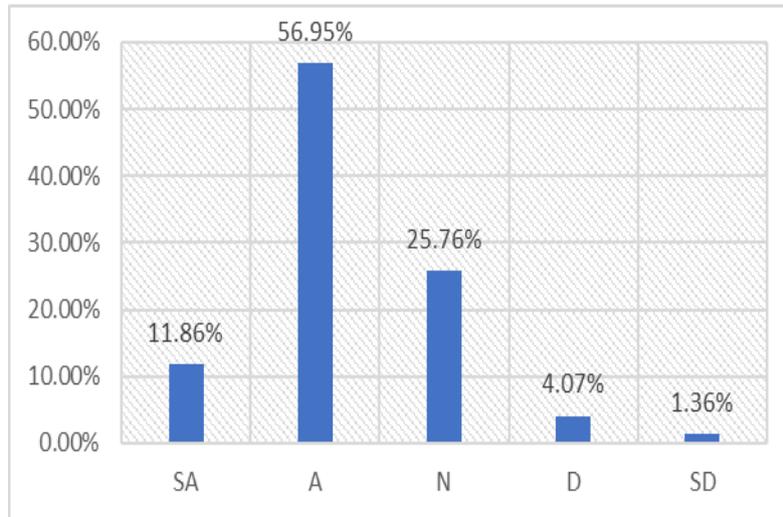

Figure 16. Mobile learning is alternative to web-based learning

In figure 17, student respondents had a neutral opinion to the statement that mobile learning is preferable than conventional or traditional learning. Only 8.14% strongly agreed and 35.56% agreed to m-learning as their preferred mode of learning.

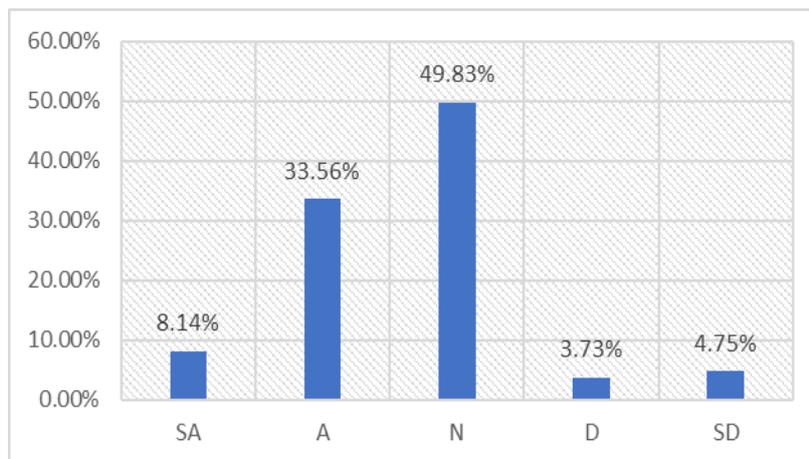

Figure 17. I prefer mobile learning than conventional learning

Overall, this study shows that more than 98% of the student respondents have their own mobile devices, carry them always and more than 15% of them have more than one 2 mobile devices which is a good indication that personal resources would not be a problem once m-learning will be implemented. In terms of students' equipment readiness, it was revealed that more than 99% of their mobile devices have applications that can be used in m-learning such as messaging apps, editing apps and different connectivity modes. With this almost perfect percentage, it was revealed that mobile devices that they own can handle some of the learning activities. In terms of students' technological skills readiness, more than 88% of the respondents have the technical know-hows in using the applications installed in their devices. Lastly, in terms of their psychological readiness, the respondents were confident in using their mobile phones as a learning tool and perceived m-learning as an alternative to web-based learning. However, these students had neutral opinions when they were asked if they prefer m-learning more than conventional learning. This goes to show that they have uncertainty about implementing m-learning as the main learning mode and still think of conventional learning to be the best way to acquire knowledge. Mobile-learning can be added to help when studying, but the traditional face to face lecture is more acceptable to them.

Now that there is a pandemic, teachers/lecturers must not rely solely on m-learning to deliver quality education given the fact that these students have means to continue their learning during home quarantine but prefer conventional learning. Students are equipped but not psychologically ready for this major change in education. Unlike the Jordanian [34] and Saudi Arabian [35] students who intends to adopt m-learning, LSPU students are not quiet there yet psychologically.

## 5. CONCLUSIONS

In consideration of the objectives of the study and results of evaluation carried out, the following conclusions were derived:

1. This study was able to assess the readiness of the students of LSPU-San Pablo City Campus towards m-learning;
2. It was able to determine that almost all of the students own mobile devices that can be used in m-learning;
3. It was able to determine that their mobile phones are fully equipped with applications that can be used in m-learning;
4. It was able to determine that the students are ready in m-learning because of their high technological skills; and
5. It was able to determine that in terms of psychological readiness, the students are quite ready.

These conclusions can be a great basis in implementing m-learning in LPSU since there is no certainty as to when do we resume to face-to-face leatning modality. Although Filipino students are ready in terms of computer/internet self-efficacy, they are not ready in terms of learner control. Sources of distraction such as social media, house chores, family duties, and work responsibilities pull the learner's focus away from his/her academic tasks, resulting in loss of productivity. Hence, it is necessary for students to maintain a healthy school-life balance even while getting educated through m-learning [36].

## 6. RECOMMENDATIONS

The following are the recommendation for enhancement of the study:

1. Consider adding other LPSU campuses such as Sta Cruz, Los Banos and Siniloan to determine the readiness of their students;
2. For further study regarding this, try other statistical treatments.
3. Psychological readiness must be considered highly because students are not fully ready yet during information gathering. Now that one year had passed, this factor must be revisited.

## 7. ACKNOWLEDGEMENT

The author is indebted to the support of College of Computer Studies and Research and Development office of Laguna State Polytechnic University, San Pablo City, Laguna, Philippines.

## AUTHOR

**Joanna E. De Torres** received her BS in Information Technology from Laguna State Polytechnic University and MIT degree from the Technological University of the Philippines. She is currently serving as Research Implementing Unit Head and a member of the Faculty of the College of Computer Studies at Laguna State Polytechnic University, Philippines. Her research interests include Performance and Reliability Analysis of Computer Software, e-Government, IT Education, and e-Learning.

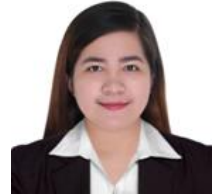